\documentstyle[sprocl]{article}

\def\Journal#1#2#3#4{{#1} {\bf #2}, #3 (#4)}

\def\be{\begin{equation}}
\def\ee{\end{equation}}
\def\bea{\begin{eqnarray}}
\def\eea{\end{eqnarray}}

\def\ket#1{| #1\rangle}

\begin{document}

\title{Beyond Boundary Conditions: General Cosmological Theories}

\author{Adrian Kent}

\address{Department of Applied Mathematics and Theoretical Physics,\\
University of Cambridge, Silver Street, Cambridge CB3 9EW, U.K.}

\maketitle\abstracts{Time-symmetric cosmological theories, in which
  the initial and final states are arranged to have similar features
  or are independently fixed, have been quite extensively discussed in
  the literature.  However, a more general and perhaps more useful
  possibility seems, surprisingly, never to have been considered.
  Quantum cosmology, like any inherently probabilistic dynamical
  theory, can be modified by conditioning on the occurrence of a 
  sequence of cosmological events fixed a priori and independently 
  of the hamiltonian and initial state.  These
  more general theories provide a natural class of alternatives which
  should eventually be useful in quantifying how well the hypothesis
  that initial conditions suffice is supported by the data.  }
  
\section{Introduction}

Standard quantum theory is explicitly time 
asymmetric.  The quantum state $\ket{\psi(t)}$ of a system
at any time $t$ is defined entirely by past events.  
It carries all the information that there is about the 
probabilities of future measurement outcomes on the system, 
but not about the probabilities of past measurement outcomes: 
to estimate the latter, we need some information about the 
state of the system prior to the measurements.  

The same is true of standard quantum cosmological models: 
the initial conditions are taken to be fixed and simple;
the probability of any given cosmological event is then determined 
by the initial conditions and (in interpretations which 
assign probabilities to sequences of events) by earlier 
events.  If standard quantum cosmology gives essentially the right picture,
then the time asymmetry of quantum theory ultimately derives from this
cosmological asymmetry.   

The hypothesis that initial causes suffice is simple, natural, 
attractive, unifying and confirmed by the many successes of 
existing cosmological theory.   Present data does not hint
that there may be any need to go beyond it.  
So it looks perverse to spend serious effort 
building cosmological theories in which 
the hypothesis does not hold --- unless these theories have some great, as 
yet unappreciated, advantage.  It may perhaps seem perverse even to mention
alternatives at all.  
Still, however sure we are of a physical principle, it's always 
interesting to ask whether it absolutely has be true and how well
it can be confirmed. 

The most obvious alternative to the standard time asymmetric
picture is a time symmetric cosmology.  Classical cosmologies with
some form of time symmetry have been discussed by several 
authors.\cite{gold}$^{\!-\,}$\cite{price} Gell-Mann and Hartle\cite{gmh} have 
recently shown how to define time-symmetric (or time-neutral) 
quantum cosmologies, in which the initial 
and final states are fixed independently of each other and of the
hamiltonian.  The hamiltonian continues to play a 
well-defined r\^ole, but the probabilities of cosmological events 
are no longer defined by the hamiltonian and initial state alone: 
they depend symmetrically on the fixed initial and final states. 

\section{Quantum Cosmologies Constrained by more than Boundary Conditions}

Time-neutral cosmologies have some surprising features,\cite{aksuper}
but they are internally consistent theories which (in principle) make
distinct empirical predictions, and have been proposed as foils
against which to test the standard understanding of time asymmetry.
However, it is hard to find examples of time-neutral cosmologies with
any theoretical appeal at all except in closed universe scenarios in
which there is recontraction to a final singularity.  It is hard
(though perhaps not impossible)
to find 
tests that would allow us to distinguish time-asymmetric and
time-neutral closed universe cosmologies.  And the data do not
currently seem to favour a closed universe.  In short, there is currently
really no very serious time-symmetric alternative to standard ideas --- 
and unless one is found, the discussion may never really affect 
practical cosmology.  Time (or CPT) symmetry in the boundary
conditions is a natural idea in its way, but to implement it 
seems to require a very large step from our current picture. 
Time-neutral cosmologies may unfortunately thus be of limited use 
as foils --- small perturbations of successful theories are generally 
more useful than large ones in this r\^ole.

Focussing on the boundary conditions, though, misses the fact that 
quantum cosmology --- or the quantum
theory of any closed system, or indeed any probabilistic dynamical
theory --- can be modified by imposing any of a large 
class of dynamical constraints. 
In essentially the same way as the predictions of a 
classical stochastic differential 
equation can be modified by restricting to the sub-ensemble of 
solutions which satisfy particular constraints at various times, a standard
quantum cosmology can be modified so that sequences of events from
some fixed sub-class of possibilities necessarily take place.  
Quite generally, given any initial state $\ket{\psi (t_0 )}$ and
hamiltonian $H$, we can, if we wish, define a theory by hypothesising
that events corresponding to the projections $P_1 , P_2 , \ldots , P_n$ 
take place at times $t_1 < t_2 < \ldots < t_n$, for any $n \leq \infty$,
and then calculate the probabilities of other events conditional 
on this hypothesis --- so long as the probability of the sequence 
of $P_i$ occurring under free evolution of $\ket{\psi(0)}$ by $H$ 
would have been non-zero. 

Now, if the present time $t > t_n$, it is true that 
the probabilities of present events could equally well
be calculated, for example, from the hypothesis that the initial state 
was 
\begin{equation} 
\label{one} 
\ket{\psi' (t_n )} = P_n \exp (i H (t_n - t_{n-1})) P_{n-1} \ldots P_1 
\exp (i H (t_1 - t_0 )) \ket{ \psi (t_ 0 )} 
\end{equation} 
at time $t_n$.
The idea here, though, is that $\ket{ \psi (t_0 )}$ and the $P_i$ 
should be relatively simple, and $\ket{\psi' (t_n )}$ rather obviously
more complicated and derivative, in the sense that its occurrence in 
a theory can only elegantly be explained via Eq.~\ref{one} and the
originally stated hypothesis.  

Instead of projections, of course, any of the more complicated notions
of quantum event discussed, for example, in the consistent histories
literature may be used.  Of these, covariant notions of event 
defined via path integrals seem fundamentally the most satisfactory
and best adapted to quantum cosmology.  
It would be possible, thus, for example, to define quantum cosmologies
in which we stipulate in advance that, when the compact 3-metric 
has volume $V_i$, the matter inhomogeneities are of scale $\delta_i$, 
for some sequence $V_1 < V_2 < \ldots$.  
There are, of course, infinitely many theories of this type, including
quite simple ones.  Since, apart from the imposed constraints,
they preserve all of standard physics, and need not differ greatly
from standard theories in their predictions, they seem ideally 
designed as foils against which to test the fundamental postulate 
that initial causes suffice. 

\section*{Acknowledgments}
This work was supported by a Royal Society University Research Fellowship.

\end{document}